On the motion of boats propelled by oars in rivers

*E94 -- De motu cymbarum remis propulsarum in fluviis*

*L. Euler*

Summary

Euler considers the following problem: A boat with a perfect rudder moves at constant speed across a stream flowing in straight fillets at assigned speeds. Assuming that the downstream velocity of the boat equals that of the river, how should the rudder be set so that the boat traverses a given path? He works out various instances, one of which gives rise to a variational problem, in detail. (From Clifford Truesdell's *An idiot's fugitive essays on science: methods, criticisms, training, circumstances*.)

Translated and Annotated[†]
by
Sylvio R. Bistafa[*]
February 2022

The calculation of the motion of boats falls into two mathematical categories, one mechanical, and the other geometrical, which differ from each other by two entirely different procedures. Indeed, those who will embark on handling this matter mechanically shall first examine the hydrostatic principles most suitable to the shape of boats; and then, by the force of the oars of the boat, he will determine the acceleration of the boat and the entire motion, both in the resting water and in a stream. The same goals apply to those wishing to use the geometrical approach, he shall first describe how the boat is directed and how it is driven by oars; and then, there are various questions that he shall address, such as how to achieve the most convenient and the quickest crossing through rivers. Indeed, in this dissertation, I have decided to develop this subject in so far as it pertains to geometry, to set aside the second part, which depends on mechanical principles. But, before I undertake this work, it is necessary that I set out and establish some hypotheses, by which this discussion shall be reduced to a purely geometrical forum.

**Hypothesis 1.**

*A boat, which is propelled by oars in quiescent water progresses with a constant motion in the direction of the spine, or in a straight line joining the prow to the stern.*

I make no assumption on this hypothesis except that which is actually observed in all the boats that are propelled by oars from rest in the water. Firstly, indeed, the rowers are assumed to row on each side of the boat with equal strength, whence the boat should be dislocated in the direction of the spine itself. Then, indeed, initially, the motion of the boat is slower, since the motion begins from rest, but soon after, it is uniform. For as soon as the boat has gained such speed, the resistance of the water equals the propulsion force, then, it will acquire neither acceleration nor retardation, and, therefore, it will be carried on with a constant motion. The inequality, however, which is present at the very beginning of the motion, may be safely neglected here, since it immediately ceases, and the beginning of the calculation can be established when the motion is actually uniform. Moreover, this same hypothesis applies when the direction of the spine is changed by the force of the rudder; for then at the same time the direction of the boat is varied, the speed remains the same. With the help of the rudder, then the boat can execute any given curve, advancing with a uniform motion; provided that the spine is always directed to the tangent of the given curve.

[†] The translator used the best of his abilities and knowledge to make this translation technically and grammatically as sound as possible. Nonetheless, interested readers are encouraged to make suggestions for corrections as they see fitting.

[*] Corresponding address: sbistafa@usp.br

**Hypothesis 2.**

*A boat being set in a river, and not driven by oars, is taken away in the same direction of the river, and it will be advanced with the same velocity that the river has.*

If the boat is moved with less velocity than the river, then it is accelerated to the velocity of the river, until the same velocity is achieved, and when that happens, the boat and the river have equal velocities, and it will be carried away in the same direction. But during the time in which the boat is still moving more slowly than the river, we here neglect, since it is very small and belongs to a mechanical work, which I do not consider here. But even if we were willing to examine the motion of the boat with the greatest rigor, of course, it would always be less than the motion of a river, because of the air resistance from that part of the boat running out of the water. But since this difference is rather small, in practice they can be safely ignored, and I will be content if these hypotheses come near at least to the truth. For my purpose is not to investigate this matter most accurately according to the laws of motion, but only according to the hypotheses for a purely geometrical approach that do not disagree from the truth very much, to which these two hypotheses are suitable.

Since, therefore, these hypotheses fit to the case where the boat is propelled by oars in still water, as well as when it is stranded in the river; it will be allowed to infer from here how a boat, driven by oars, ought to advance in the river. This is indeed the case where the boat will be moved by a composed motion, which arises from two situations, certainly one in which it moves if the water is at rest, and the other in which it moves if the oars were missing. Therefore, a composition of motion is invoked, and our entire work may be completed by geometry alone; and on that account, I will proceed to solve the following problems.

**Problem I.**

*Given the velocity of the river in each location, and the direction which the spine of the boat has everywhere, to find the curve which the boat will describe in the river.*

**Solution.**

Line $AB$ (Table III, Fig. 1) traverses the course of the river normally, being $AMC$ the desired curve, in which the boat or rather its center of gravity $M$ is moved; from this curve, we will apply $MP$ in the direction of the straight line $AB$ such as an axis in the direction of a river. Let us consider that the boat is found at $M$, where the direction of the spine at this location is $aMb$, which, with the direction of the river $PM$, constitute the angle $P\widehat{M}b$, being $m$ and $n$ the sine and cosine of this angle, such that $m^2 + n^2 = 1$. Set $c$ as the velocity of the boat when propelled by oars in still water, with which it moves uniformly according to the direction of the spine; being $u$, indeed, the velocity in which the river advances in $M$ in the direction $Mq$; being this velocity somehow variable. Truly, however, the desired velocity with which the boat or rather its center of gravity $M$ is actually moved along the curve $AMC$ is set as $= v$.

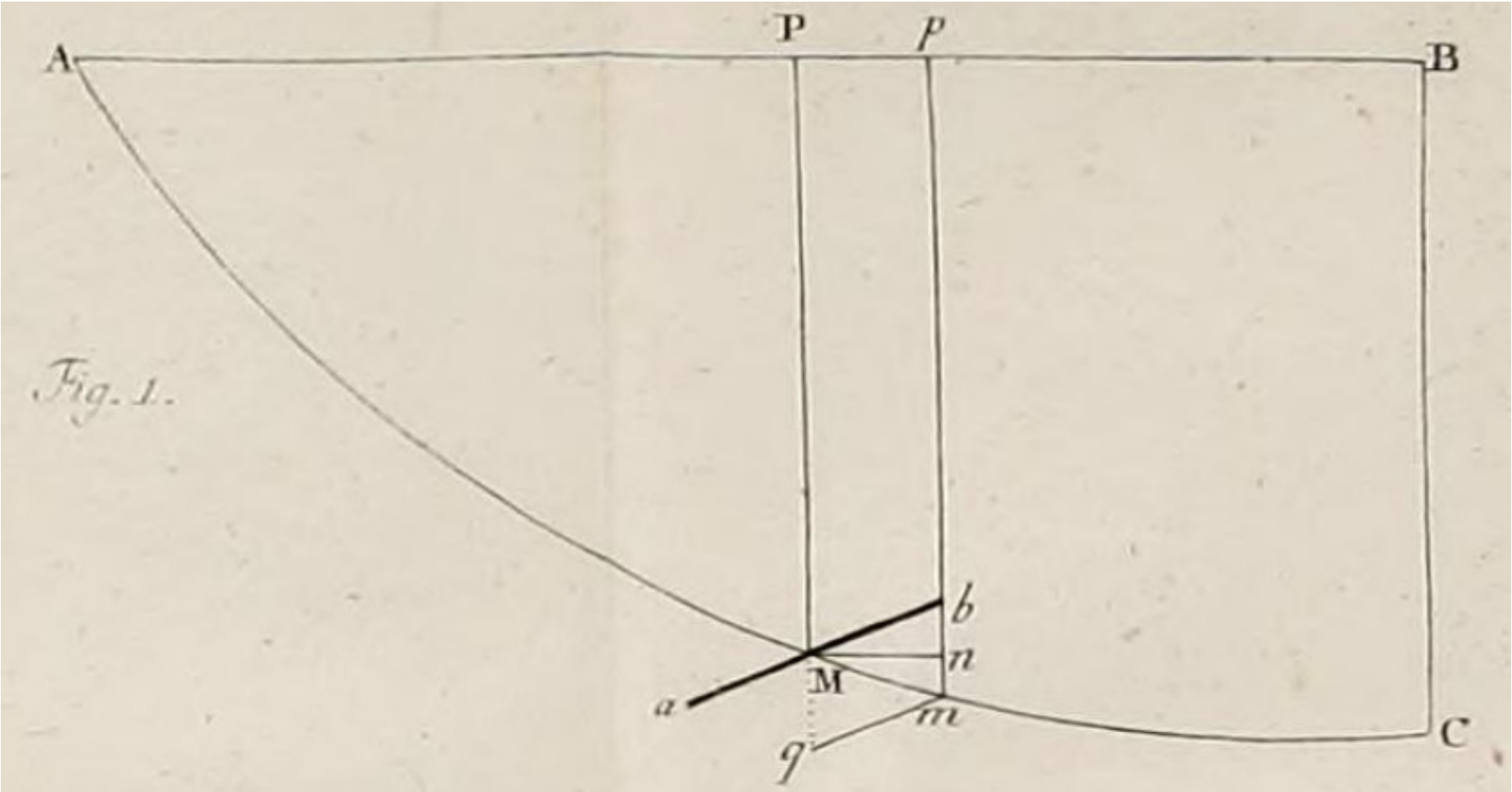

Table III, Figure 1

Set now the abscissa $AP = x$; ordinate $PM = y$, and considering $AM = s$, may be guided the ordinate $pm$ in the proximity, such that $Pp = Mn = dx$; $mn = dy$ and also $Mm = ds$. If now the velocity of the river ceases and the boat is driven only by the oars, it advances in the direction of the spine $ab$ with the velocity $c$, and in a certain time the center of gravity $M$ reaches point $b$ (as per hyp. 1). But if the force of the oars ceases, the boat is conducted only by the course of the river, and then $M$ is driven in the direction $PM$, with the velocity $= u$, which, during the same element of time, it reaches $q$ (as per hyp. 2) so that it will be $Mb\!:\!Mq = c\!:\!u$. Therefore, if the boat is pushed on by both forces, namely jointly by the oars and the river, then it is necessary that it advances along the diagonal $Mm$ of the parallelogram $Mbmq$, with a velocity proportional to this diagonal, to such an extent that $Mm\!:\!Mb = v\!:\!c$ or $Mm\!:\!Mq = v\!:\!u$. Because $\sin P\widehat{M}b = \cos b\widehat{M}n = m$, and $\cos P\widehat{M}b = \sin b\widehat{M}n = n$; then, $\tan b\widehat{M}n = \frac{n}{m} = \frac{bn}{Mn}$; whence $bn = \frac{ndx}{m}$ and $bm = Mq = dy + \frac{ndx}{m}$. But, since $\frac{Mn}{Mb} = m$, then, $Mb = \frac{dx}{m}$. Since, indeed, $Mb\!:\!Mq = c\!:\!u$, then, $dx\!:\!mdy + ndx = c\!:\!u$, giving the equation of the desired curve as $udx = cmdy + cndx$; or $dy = \frac{dx(u-cn)}{cm}$. Moreover, the actual velocity $v$ of the boat, with which it is moved along this curve, it will be known from the ratio $ds\!:\!\frac{dx}{m} = v\!:\!c$, whence $v = \frac{cmds}{dx}$. *Q.E.I.*

*Corollary 1.* Since $ds = \sqrt{dx^2 + dy^2}$, then, substituting the expression just found $\frac{dx(u-cn)}{cm}$ for $dy$, results in $ds = \sqrt{c^2 - 2cnu + u^2}$.

*Corollary 2.* From the given velocity of the boat at individual locations, the time that the boat takes to complete the curve $AM$ will be known: certainly this time will be $= \int \frac{ds}{v} = \int \frac{dx}{cm}$.

*Corollary 3.* If, therefore, the direction of the boat or the angle $P\widehat{M}b$ is determined by the abscissa $AP$ alone, also the time at which the corresponding arc of the given abscissa is completed will be defined by the abscissa alone, neither this mutability will depend on velocity of the river.

*Corollary 4.* It is understood from the equation of the curve $AMC$, just found to be given by $dy = \frac{dx(u-cn)}{cm}$, that where $u > cn$, there the boat descends in the river, where, indeed, $u < cn$, there it ascends; finally, in the location where $u = cn$, there the boat stands most apart from the line $AB$, and in that place, the tangent of the described curve will be parallel to the line $AB$.

*Corollary 5.* If the actual motion of the boat through $Mm$ is decomposed into a crossing motion in the direction of the line $AB$, and into an ascending or descending motion, whose direction is a priori normal and in the direction that coincides with the course of the river, then the velocity of the crossing motion $= cm$; and, indeed, the velocity of the descending motion $= \frac{cmdy}{dx} = u - cn$.

*Corollary 6.* Therefore, the boat fastest crossing motion occurs when $m = 1$, that is, when the spine of the boat is always directed to the direction normal to the river. Moreover, in this case, the time to reach the bank of the river at a distance $AP = x$ will be $= \frac{x}{c}$.

*Corollary 7.* But, if the boat is directed against the direction of the course of this river, then $m = 0$,and $n = 1$, whence the crossing ceases, and the boat ascends or descends, just as $u$ has been smaller or greater than $c$, then, obviously, the velocity with which it ascends $= c - u$.

*Corollary 8.* But if the prow of the boat is directed downwards, then $n = -1$, and the boat descends directly straight with the velocity $c + u$, that is, with the velocity of the river combined with the velocity of the boat, which it has when moved in still water.

*Scholion.* In the solution of this problem I have placed the prow of the boat $b$ so that the angle $P\widehat{M}b$, which the direction of the direction of the boat makes with the flowing river is acute; but the same solution is equally applicable at obtuse angles. Thus, if the angle $P\widehat{M}b$ would be obtuse, then its cosine $n$ must be taken as negative, and the same equation which I found will be valid for this case. However, since this solution is very general, it will be advantageous to have a clearer understanding of some particular cases, to which it seemed appropriate to add the following problems.

**Problem II.**

*If the boat $ab$ always keeps the same angle $P\widehat{M}b$ with the direction $PM$ of the river, find the curve $AMC$ in which the boat will be moving.*

**Solution.**

Set $AEFB$ (Table III, Fig. 2) as the channel of the river, and $AB$ the line that traverses the river normally, which was drawn from point $A$ where the boat departs. Set the velocity with which the boat advances $= c$, which is expressed by the line segment $AD = c$; in particular, the curve $AQB$ exhibits the velocity of the river along its width, such that its ordinate $PQ$ indicates the velocity with which the portion of the river $PMmp$ slides; therefore, setting $AP = x$, then $PQ = u$. Furthermore, set $AMC$ as the desired curve, in which the center of gravity $M$ of the boat $ab$ is moved along, and also its ordinate $PM = y$: indeed, the true velocity which the boat traverses the element $Mm$ is $= v$; finally, the sine of the angle $P\widehat{M}b$ is $= m$ and the cosine is $= n$, considering the whole sine $= 1$, which, by hypothesis, are constant. Then, from these, $dy = \frac{dx(u-cn)}{cm}$, and also $y = \int \frac{udx}{cm} - \frac{nx}{m}$: whence $PM = \frac{area\ APQ}{m \cdot AD} - \frac{n}{m}AP$; from this equation, then follows an easy construction of the desired curve $AMC$ from the quadrature of the curve $AQB$. Specifically, the time that the boat takes to complete the arc $AM$ will be $= \int \frac{dx}{cm} = \frac{x}{cm} = \frac{AP}{m \cdot AD}$.

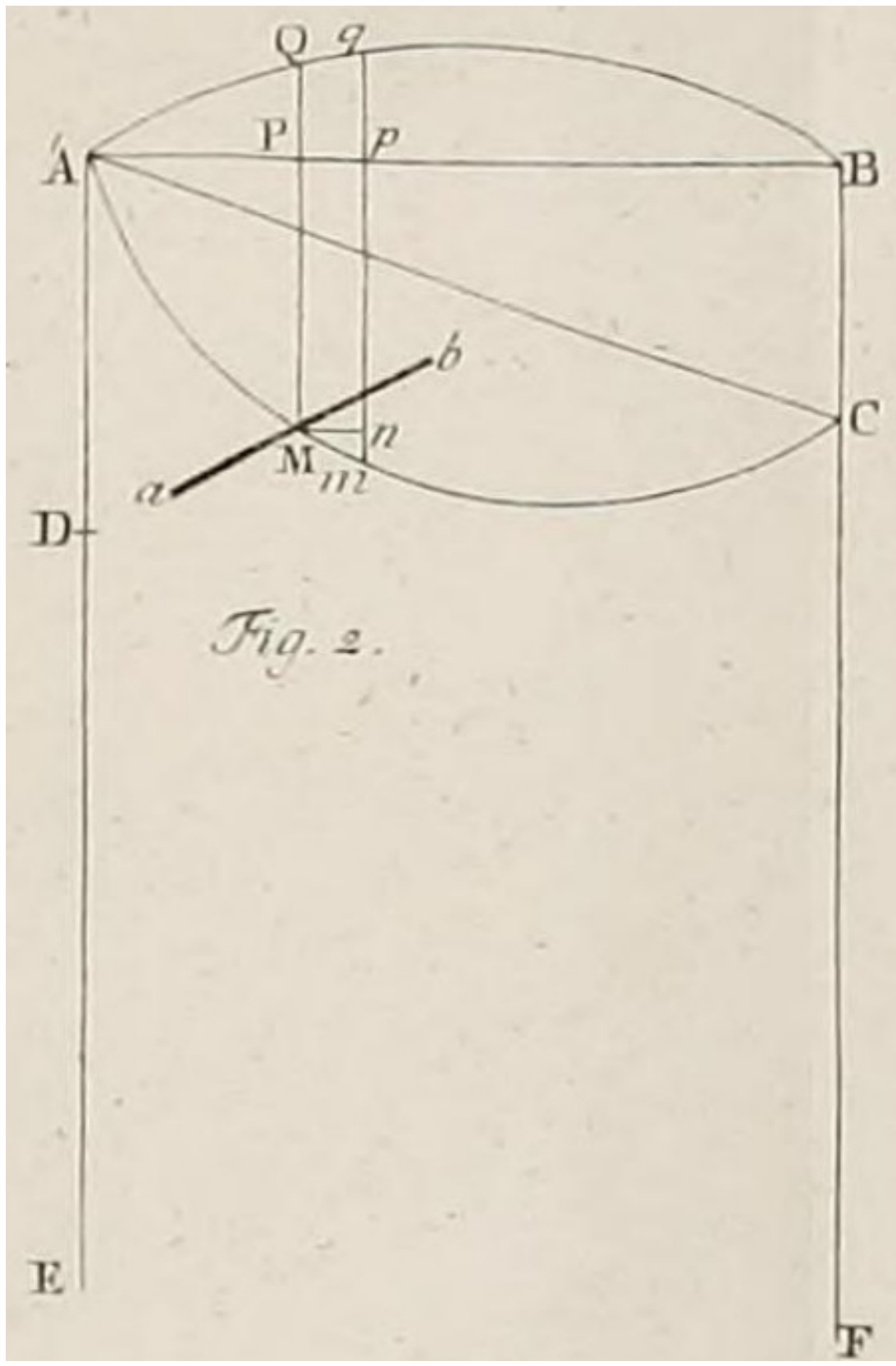

Table III, Figure 2

*Corollary 1.* Since $dy = \frac{dx(u-cn)}{cm}$, then, by considering $dx$ as constant, we will have that $d^2y = \frac{dxdu}{cm}$. Therefore, the curve $AMC$ will have a point of inflection where $du = 0$, that is, where the velocity of the river is maximum.

*Corollary 2.* The tangent of the angle $A\widehat{M}P$, which the curve described by the boat makes with the course $PM$ of the river is $= \frac{dx}{dy} = \frac{cm}{(u-cn)}$. Therefore, where the velocity of the river vanishes, there the direction of the spine is tangent to the curve.

*Corollary 3.* If $C$ is a point where the boat lands, which opposes the river bank, then $BC = \frac{area\ AQBA}{m \cdot AD} - \frac{n}{m}AB$. Therefore, if the area $AQBA = nAB \cdot AD$, then the boat lands at point $B$.

*Corollary 4.* If we set $AB = a$ and the area $AQBA = ab$, such that $BC = f$, then, $f = \frac{ab}{mc} - \frac{na}{m}$, then, from this equation, we will have that $mcf = ab - nac$, and $m^2c^2f^2 = c^2f^2 - n^2c^2f^2 = a^2b^2 -$ $2na^2bc + n^2a^2c^2$, whence $n = \frac{a^2b(+/-)f\sqrt{a^2c^2+c^2f^2-a^2b^2}}{(a^2+f^2)c}$ , $m = \frac{abf(-/+)a\sqrt{a^2c^2+c^2f^2-a^2b^2}}{(a^2+f^2)c}$; and $\frac{m}{n} = \frac{ac^2f(+/-)\sqrt{a^2c^2+c^2f^2-a^2b^2}}{a^2b^2-c^2f^2}$.[a]

*Corollary 5.* Therefore, if the boat ought to land at a given point $C$, the tangent of the angle $PM\widehat{M}b$ which the boat should make, at all times, with the direction of the course of the river will be $= \frac{ac^2f\mp ab\sqrt{a^2c^2+c^2f^2-a^2b^2}}{a^2b^2-c^2f^2}$. Then, unless $c\sqrt{(a^2+f^2)} > ab$, that is, unless $AD > \frac{area\ AQBA}{AC}$, the boat cannot land at point $C$[b].

*Corollary 6.* Since $m$ cannot have a negative value, for otherwise the boat will not come to the opposite bank, the only way that the boat can take to reach point $C$ is that for which $bf < \sqrt{a^2c^2+c^2f^2-a^2b^2}$ or $b < c$.[c]

*Corollary 7.* But, if $bf > \sqrt{a^2c^2+c^2f^2-a^2b^2}$ or $b > c$, then the boat will be able to reach point $C$ in two ways, because of the double positive value of $m$. But there must also be $f > \frac{a}{c}\sqrt{b^2-c^2}$; for $m$ not to be imaginary.

---

[a] Note that signs before the square-root of $n$ and $m$ are reversed in the ratio $m/n$

$$\begin{aligned}\frac{m_1}{n_1} = \frac{abf + a\sqrt{a^2c^2+c^2f^2-a^2b^2}}{a^2b - f\sqrt{a^2c^2+c^2f^2-a^2b^2}} &= \frac{\left(abf + a\sqrt{a^2c^2+c^2f^2-a^2b^2}\right)\cdot\left(a^2b + f\sqrt{a^2c^2+c^2f^2-a^2b^2}\right)}{a^4b^2 - f^2(a^2c^2+c^2f^2-a^2b^2)} \\ &= \frac{a^3b^2f + \left[ab(a^2+f^2)\cdot\sqrt{a^2c^2+c^2f^2-a^2b^2}\right] + af(a^2c^2+c^2f^2-a^2b^2)}{(a^2+f^2)(a^2b^2-c^2f^2)} \\ &= \frac{a^3b^2f + a^3c^2f + ac^2f^3 - a^3b^2f + \left[ab(a^2+f^2)\cdot\sqrt{a^2c^2+c^2f^2-a^2b^2}\right]}{(a^2+f^2)(a^2b^2-c^2f^2)} \\ &= \frac{(a^2+f^2)acf + \left[ab(a^2+f^2)\cdot\sqrt{a^2c^2+c^2f^2-a^2b^2}\right]}{(a^2+f^2)(a^2b^2-c^2f^2)} = \frac{acf + ab\sqrt{a^2c^2+c^2f^2-a^2b^2}}{a^2b^2-c^2f^2} \\ &= \frac{acf + ab\sqrt{a^2c^2+c^2f^2-a^2b^2}}{a^2b^2-c^2f^2}\end{aligned}$$

$$\begin{aligned}\frac{m_2}{n_2} = \frac{abf - a\sqrt{a^2c^2+c^2f^2-a^2b^2}}{a^2b + f\sqrt{a^2c^2+c^2f^2-a^2b^2}} &= \frac{\left(abf - a\sqrt{a^2c^2+c^2f^2-a^2b^2}\right)\cdot\left(a^2b - f\sqrt{a^2c^2+c^2f^2-a^2b^2}\right)}{a^4b^2 - f^2(a^2c^2+c^2f^2-a^2b^2)} \\ &= \frac{a^3b^2f - \left[ab(a^2+f^2)\cdot\sqrt{a^2c^2+c^2f^2-a^2b^2}\right] + af(a^2c^2+c^2f^2-a^2b^2)}{(a^2+f^2)(a^2b^2-c^2f^2)} \\ &= \frac{a^3b^2f + a^3c^2f + ac^2f^3 - a^3b^2f - \left[ab(a^2+f^2)\cdot\sqrt{a^2c^2+c^2f^2-a^2b^2}\right]}{(a^2+f^2)(a^2b^2-c^2f^2)} \\ &= \frac{(a^2+f^2)acf - \left[ab(a^2+f^2)\cdot\sqrt{a^2c^2+c^2f^2-a^2b^2}\right]}{(a^2+f^2)(a^2b^2-c^2f^2)} = \frac{acf - ab\sqrt{a^2c^2+c^2f^2-a^2b^2}}{a^2b^2-c^2f^2} \\ &= \frac{acf - ab\sqrt{a^2c^2+c^2f^2-a^2b^2}}{a^2b^2-c^2f^2}\end{aligned}$$

[b] $a^2+f^2 = AB^2 + BC^2 = AC^2, (see\ Table\ III, Figure\ 2)$; $c\sqrt{(a^2+f^2)} > ab \Rightarrow AD\cdot AC > area\ AQBA \Rightarrow AD > \frac{area\ AQBA}{AC}$.

[c] For $b = c$, then $\sqrt{a^2c^2+c^2f^2-a^2b^2} = cf$, hence $bf < \sqrt{a^2c^2+c^2f^2-a^2b^2}$ amounts to $b < c$. In this case, in the numerator of the expression for $m$, $abf < acf$. Since $m$ cannot be negative, the plus sign before the radical applies, and there is only one way to reach point $C$, that for which $m = \frac{abf+a\sqrt{a^2c^2+c^2f^2-a^2b^2}}{(a^2+f^2)c}$.

*Corollary 8.* However, in these cases, in which the boat reaches $C$ from two angular directions, half the sum of these two angles is equal to the angle $B\hat{A}C$, drawn by the chord $AC$ [d]. Indeed, from half the difference of these two angles the sine $= \frac{ab}{c\sqrt{(a^2+f^2)}}$ and the cosine $= \frac{2ab\sqrt{a^2c^2+c^2f^2-a^2b^2}}{c^2(a^2+f^2)}$. Whence, from the half the sum and from half the difference each angle is found easily enough.

*Corollary 9.* Therefore, when the angle which is half the difference is smaller than the angle $B\hat{A}C$, then the point $C$ can be reached by a double angle. Also, the boat calls $C$ sooner, by following the greater angle, or rather the one whose sinus is greater; the time that the boat takes to cross the river is $= \frac{AB}{mAD}$, where $m$ is the sine of the angle which the direction of the boat makes with the course of the river.

*Scholion.* From the equation $PM = \frac{area\ APQ}{m \cdot AD} - \frac{n}{m}A$, which I found for the curve described by the boat, this curve is fairly easily constructed as follows (Table IV, Fig. 1). Indeed, given the curve $ALQB$, which expresses the velocity of the river, may be drawn straight the line $GAH$ through $A$, parallel to the direction of the boat at each instant, and from $D$, the perpendicular $DG$ to this line is dawn such that $DG = m \cdot AD$ and $PR = \frac{n}{m}AP$. To this end, the curve described is found by taking from any point M $PM = \frac{area\ APQ}{m \cdot AD} - PR$. Indeed, the time that the boat takes to reach $M$ from $A$ is expressed by $\frac{AP}{DG}$. However, it should be noted that if the curve $ALB$ passes through the points $A$ and $B$, which generally takes place in all the rivers, seeing that the flow is faster in the middle and slower towards the banks, then not only is $AH$ the tangent of the curve at $A$, but also the tangent at $C$ is parallel to $AH$ itself.

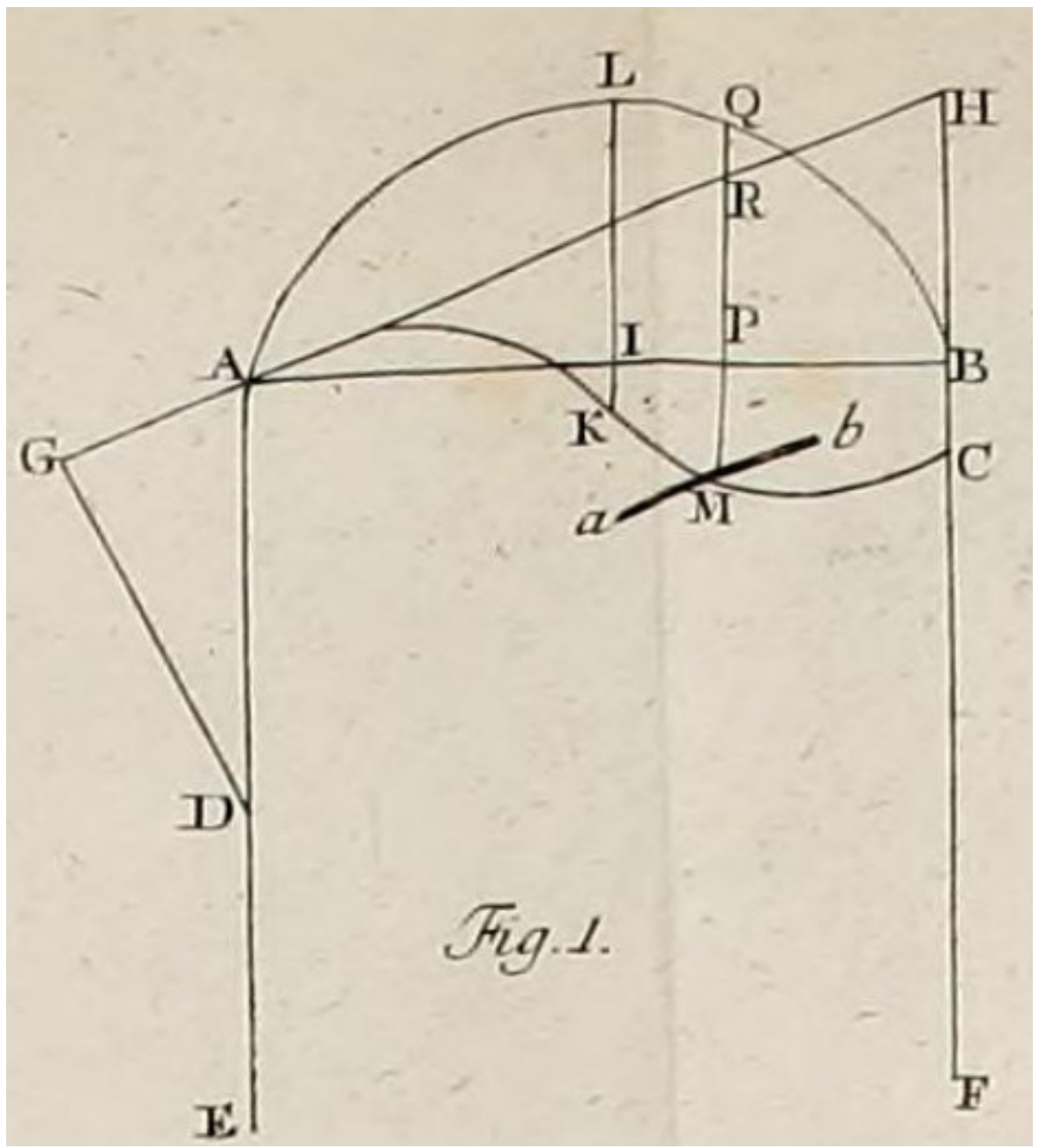

Table IV, Figure 1

[d] It is not clear where this construction comes from.

Moreover, if a river at equal distances from both banks has equal velocities, so that the curve $ALB$ has the diameter $LIK$ at equal distances between the banks $AE$ and $BF$; then the curve described by $AKC$ will have two equal parts $AK$ on one side and $KC$ on the other side of $K$; in $K$, however, it will have a point of inflection in the opposite direction, as is clearly seen from above.

**Problem III.**

*If a boat, crossing a river is always directed toward a fixed point $H$, define the curve $AMC$ (Table III, Figure 3) which the boat, departing from $A$, will describe in the river.*

**Solution.**

Be as before the curve $AQB$ the scale of the velocities of the river, and $AD$ the velocity $c$ which the boat moves in still water; and also $AP = x$, $PM = y$; $PQ = u$, $\sin P\widehat{M}b = m$, $\cos P\widehat{M}b = n$. Also, the width $AB$ of the river $= a$, $BG = g$ and $GH = b$; from the fixed point $H$, the normal $HGK$ to $PM$ is drawn. Then, $HK = a + b - x$, and $KM = y - g$. But, because the direction of the boat points to $H$, then, $\frac{a+b-x}{y-g}$ is the tangent of the angle $P\widehat{M}b$ and, for that reason $= \frac{m}{n}$, whence, $m = \frac{a+b-x}{\sqrt{(y-g)^2+(a+b-x)^2}}$. But, since the equation for the desired curve as found above is $dy = \frac{udx}{cm} - \frac{ndx}{m}$, then, for our case $dy = \frac{udx\sqrt{(y-g)^2+(a+b-x)^2}-c(y-g)dx}{c(a+b-x)}$. Indeed, the time that the boat takes to reach $M$ from $A$ will be $= \int \frac{dx\sqrt{(y-g)^2+(a+b-x)^2}}{c(a+b-x)}$. *Q.E.I.*

*Corollary 1.* However, in the equation just found, the variables $y, x$ and $u$ which depend on $x$, are interchangeable, yet, if it is considered that $y - g = (a + b - x)z$ they will be separated from each other, giving, indeed, this equation $\frac{cdz}{\sqrt{1+z^2}} = \frac{udx}{a+b-x}$.[e]

*Corollary 2.* Because $u$ depends on $x$, let us set $\int \frac{udx}{a+b-x} = \log_c X$, which is an acceptable integral that vanishes for $x = 0$. Henceforth, we will have that $c \cdot \log_c \left(z + \sqrt{(1 + z^2)}\right) = \log_c X + Const.$, that for the determination of the constant it is set $x = 0$, giving $z = \frac{-g}{a+b}$, and hence $Const. = c \cdot \log_c \left(\frac{-g+\sqrt{g^2+(a+b)^2}}{a+b}\right)$.

*Corollary 3.* Then, since $\int \frac{udx}{a+b-x} = \log_c X$ or $X = c^{\int \frac{udx}{a+b-x}}$, we have the following integral equation for the desired curve[f], $X^{1/c} = \frac{(a+b)z+(a+b)\sqrt{1+z^2}}{-g+\sqrt{g^2+(a+b)^2}}$ or [g] $\frac{X^{1/c}(a+b-x)}{a+b} = \frac{y-g+\sqrt{(y-g)^2+(a+b-x)^2}}{-g+\sqrt{g^2+(a+b)^2}}$.

---

[e] *Since* $z = \frac{y-g}{(a+b-x)}$, *then* $dz = \frac{(y-g)dx}{(a+b-x)^2} + \frac{dy}{(a+b-x)}$. We also have that $dy = \frac{udx\sqrt{(y-g)^2+(a+b-x)^2}-c(y-g)dx}{c(a+b-x)}$; then, $dz = \frac{dx}{(a+b-x)^2} + \frac{udx\sqrt{(y-g)^2+(a+b-x)^2}-c(y-g)dx}{c(a+b-x)^2}$. Also, since $z = \frac{n}{m} = \frac{y-g}{a+b-x}$, then, indeed, $\frac{cdz}{\sqrt{1+z^2}} = \frac{c(y-g)dx}{(a+b-x)\sqrt{(a+b-x)^2+(y-g)^2}} + \frac{udx}{(a+b-x)} - \frac{c(y-g)dx}{(a+b-x)\sqrt{(a+b-x)^2+(y-g)^2}} = \frac{udx}{(a+b-x)}$.

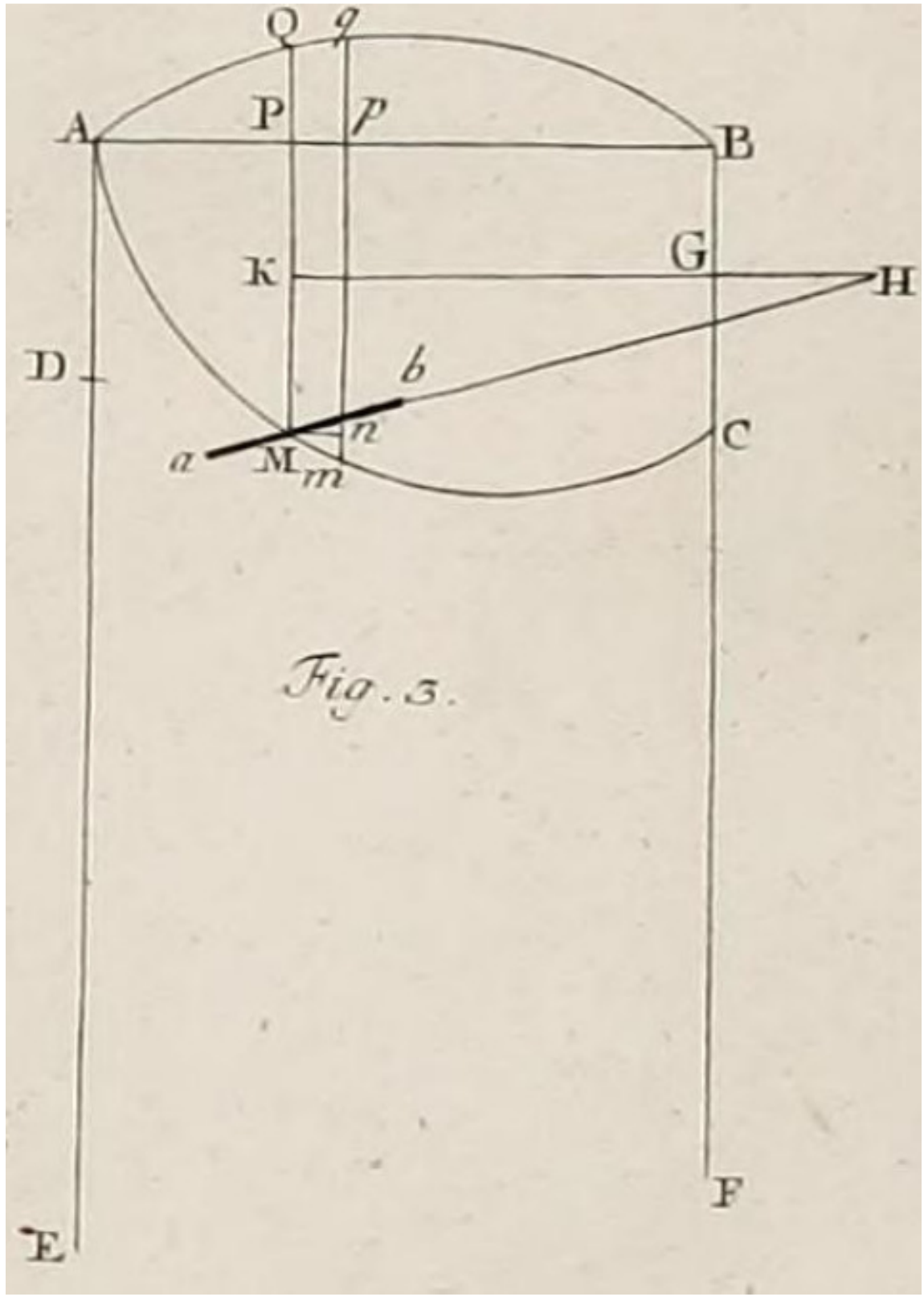

Table III, Figure 3

*Corollary 4.* From the equations just found for $y$ or $z$ as a function of $x$, the time that the boat takes to reach $M$ from $A$ is [h] $= \int \frac{dx\sqrt{(y-g)^2+(a+b-x)^2}}{c(a+b-x)} = \int \frac{dx\sqrt{1+z^2}}{c}$, which can be found by quadrature.

*Corollary 5.* If the point $H$ falls into point $G$ or into the river bank itself, then $b = 0$, and we will have that $\frac{X^{1/c}(a-x)}{a+b} = \frac{y-g+\sqrt{(y-g)^2+(a-x)^2}}{-g+\sqrt{(g^2+a^2)}}$. If now we set $x = a$, which then gives $y = BC = f$, meaning that the point $C$ where the boat lands on bank $BF$ is known, it will be found that $\frac{2f-2g}{-g+\sqrt{(g^2+a^2)}}$, then, unless in this case, the quantity $X^{1/c}$ by chance becomes infinitely large, which when applied in $a - x = 0$ gives a finite quantity.

---

[f] $c \cdot \log_c\left(z+\sqrt{1+z^2}\right) = \log_c X + Const. \Rightarrow \log_c\left(z+\sqrt{1+z^2}\right) = \log_c X^{1/c} + \log_c\left(\frac{-g+\sqrt{g^2+(a+b)^2}}{a+b}\right) \Rightarrow$ $\log_c\left(z+\sqrt{1+z^2}\right) = \log_c\left[X^{1/c}\cdot\left(\frac{-g+\sqrt{g^2+(a+b)^2}}{a+b}\right)\right] \Rightarrow X^{1/c} = \frac{(a+b)z+(a+b)\sqrt{1+z^2}}{-g+\sqrt{g^2+(a+b)^2}}$

[g] We have that $\frac{a+b-x}{a+b} = \frac{y-g}{z(a+b)}$, then,

$$\frac{X^{1/c}(a+b-x)}{a+b} = \frac{y-g+\frac{(y-g)}{z}\sqrt{1+z^2}}{-g+\sqrt{g^2+(a+b)^2}} = \frac{y-g+\frac{(y-g)}{z}\sqrt{1+\frac{(y-g)^2}{(a+b-x)^2}}}{-g+\sqrt{g^2+(a+b)^2}} = \frac{y-g+\frac{(y-g)}{(a+b-x)z}\sqrt{(y-g)^2+(a+b-x)^2}}{-g+\sqrt{g^2+(a+b)^2}} = \frac{y-g+\sqrt{(y-g)^2+(a+b-x)^2}}{-g+\sqrt{g^2+(a+b)^2}}$$

[h] $\sqrt{1+z^2} = \frac{\sqrt{(y-g)^2+(a+b-x)^2}}{(a+b-x)}$

*Corollary 6.* Therefore, if I make $x = a$, the quantity $\frac{X^{1/c}(a-x)}{a+b}$ vanishes, giving $f = g$ or $BC = BG$. In these cases, therefore, the boat calls the point $G$ to which it was always directed.

*Corollary 7.* But from the very nature of the matter, it is understood if the edge of the curve $AB$ applied at $B$ were either $= 0$ or less than $c$, that is, if a river is carried along the bank $BF$ with less velocity than a boat is propelled, then, the boat will be always be put ashore at point $G$, if, indeed, $H$ falls in $G$. For if it was moving toward another point, then, it would continue to be moved by directing the motion toward $G$, until it arrives at $G$.

**Scholion.**

In order that the motion of this boat may be more clearly known, I will present an example by which we may fully disclose the equation just found. Let a river (Table IV, Figure 2), certainly having the same velocity everywhere, by which the scale of the velocities may be considered a straight line $ab$ parallel to the axis $AB$, and may the boat be constantly directed toward the fixed point $G$, situated at the opposite bank, such that $BG = g$.

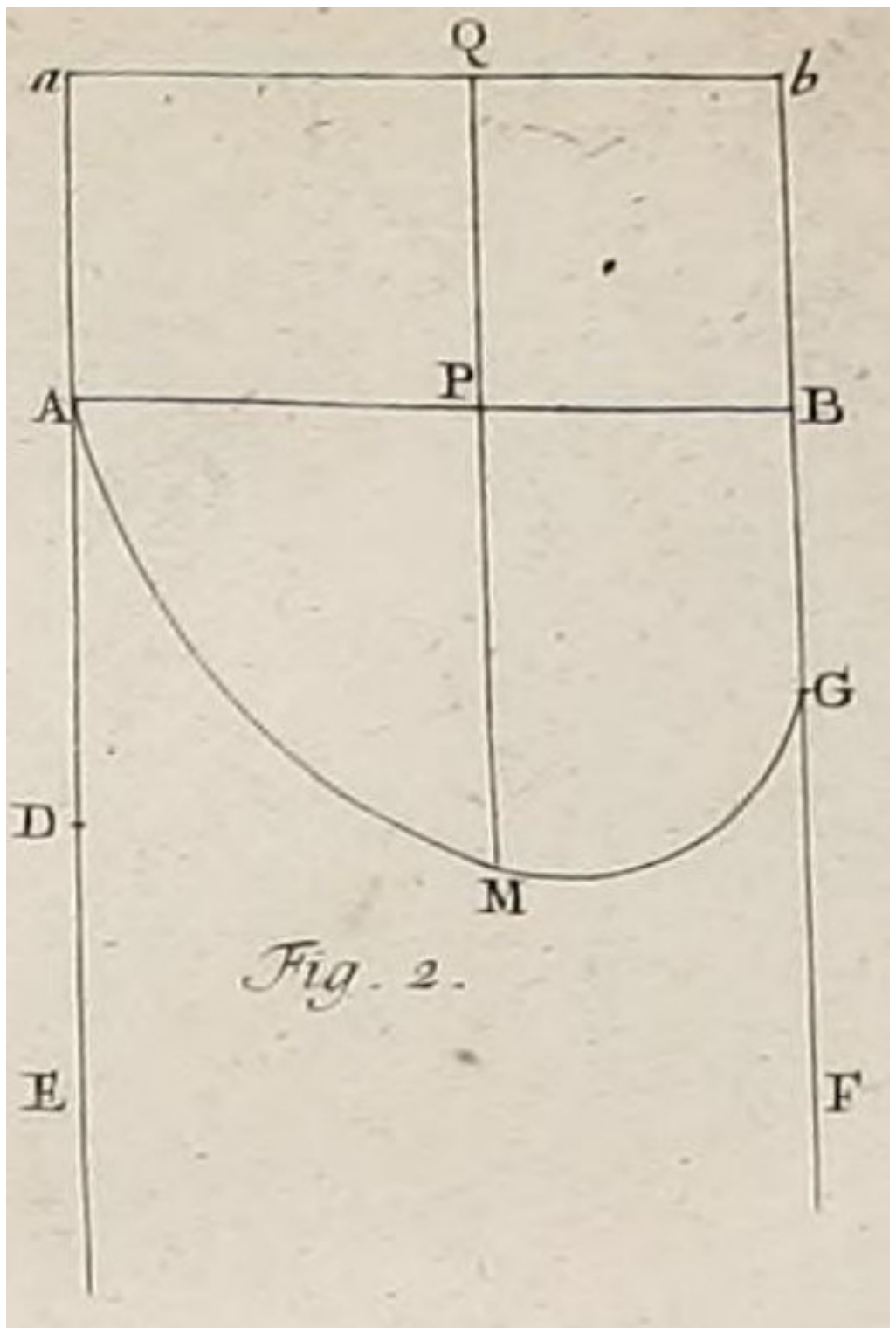

Table IV, Figure 2

Thus, $u$ will become a constant quantity, which is $=\propto c$. Then, for this case the following will hold

$$log_c X =\propto c \int \frac{dx}{a-x} =\propto c\ log_c \frac{a}{a-x}\ or\ X = \frac{a^{\propto c}}{(a-x)^{\propto c}}.$$

When this expression is substituted into the equation just found it will give

$$\frac{a^{\propto-1}}{(a-x)^{\propto-1}}=\frac{y-g+\sqrt{(y-g)^2+(a-x)^2}}{-g+\sqrt{(g^2+a^2)}}.$$

Hence, it is clear that if $\propto<1$ [$u<c$], then the boat will be landing at the point $G$; however, if $\propto>1$ [$u>c$], then the boat cannot reach the bank at all. Also, the case for which $\propto=1$, or $u=c$, we will have that $\sqrt{(g^2+a^2)}=y+\sqrt{(y-g)^2+(a-x)^2}$ [here, Euler wrongly writes $\sqrt{(g^2a^2)}=y+\sqrt{(y-g)^2+(a-x)^2}$] or $x^2-2ax=2gy-2y\sqrt{(a^2+g^2)}$, which is the equation for a parabola with axis $BF$ and the vertex in $F$, where $BF=\frac{a^2}{2\sqrt{(a^2+g^2)}-2g}$, whose parameter is $2\sqrt{(a^2+g^2)}-2g$; so that, then, the focus of this parabola falls right at point $G$. Therefore, the parabola will be described from the given focus $G$, the position of the axis $FB$ and the point $A$ where the parabola must pass through.

**Problem 4.**

*Given the scale $AQB$ of velocities of the river (Table IV, Figure 3), find the direction of the boat in every location such that it describes a given curve $AMC$.*

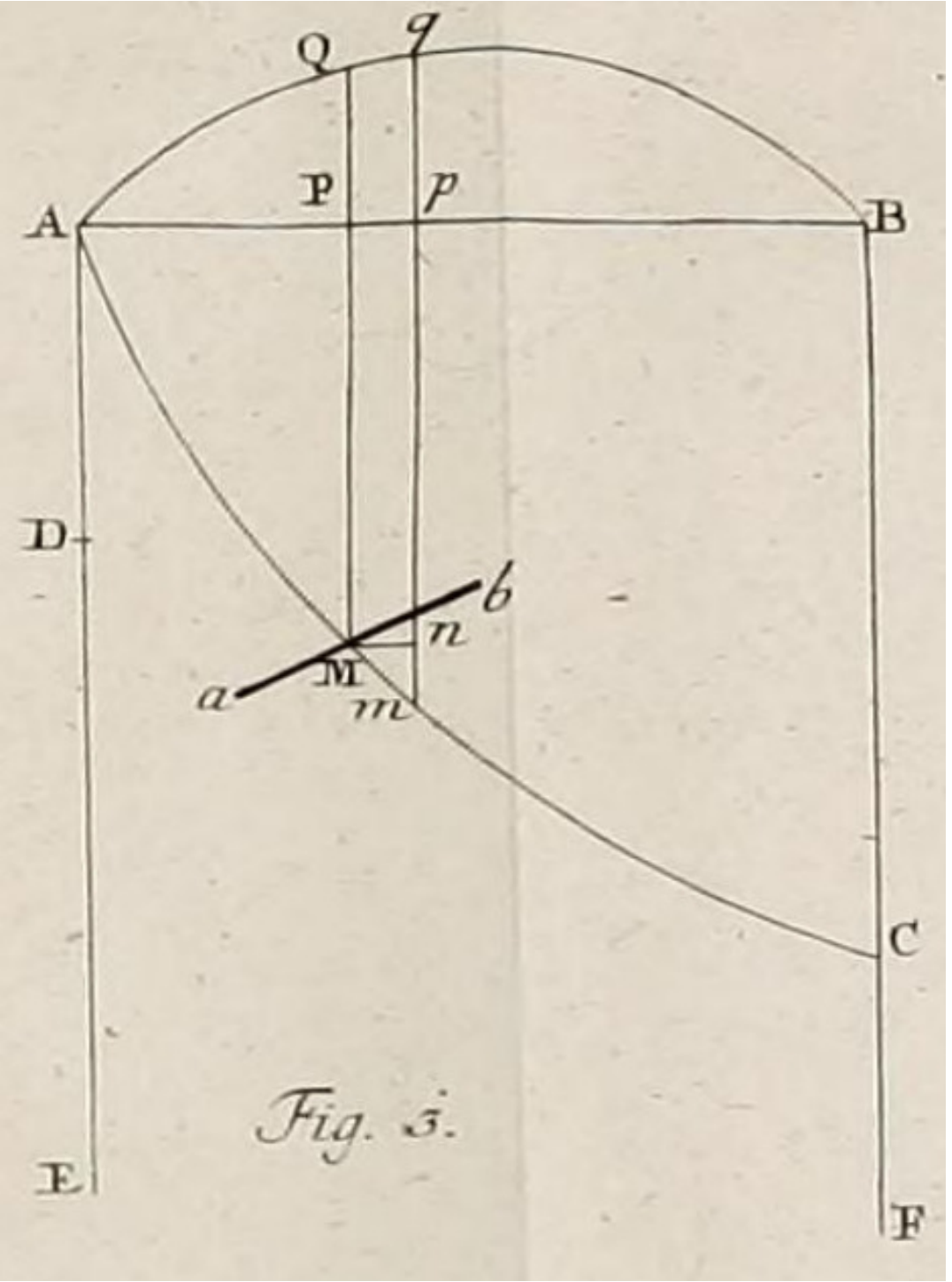

Table IV, Figure 3

**Solution.**

Just as the line $PQ$ of the curve $AQB$ indicates the velocity of the river in each location $P$, so let $AD$ be the velocity which the boat advances in the direction of the spine in still water. Thus, set $AD=c$; $AP=x$; $PQ=u$; and be set in the curve to be described by the boat $PM=y$; and the arc $AM=s$, such that $ds=\sqrt{dx^2+dy^2}$. Consider now that $ab$ is the desired direction which the boat should attain at every point $M$, such that it is moved along a given curve $AMC$; and that the sinus of the angle $P\widehat{M}b$ is set as $=m$ and its cosine $=n$, considering that the whole sine $=1$. There exist, therefore, an equation between $x$ and $u$, and also an equation between $x$ and $y$, from which it is necessary to define either $m$ or $n$. However, we found earlier in the first problem the equation $udx - cmdy = cndx = c\sqrt{1-m^2}dx$, since $n = \sqrt{1-m^2}$. By taking the square of this equation we have that $u^2dx^2 - 2cmudxdy + c^2m^2dy^2 = c^2dx^2 - c^2m^2dx^2$, which can be transformed into $c^2m^2 = \frac{2cmudxdy + c^2dx^2 - u^2dx^2}{dx^2+dy^2}$, and by extracting the two roots of the former equation results in $cm = \frac{udxdy \mp dx\sqrt{c^2dx^2+c^2dy^2-u^2dx^2}}{dx^2+dy^2}$ , [and from $udx - cmdy = cndx$], we have that $cn = \frac{udx^2 \mp dy\sqrt{c^2dx^2+c^2dy^2-u^2dx^2}}{dx^2+dy^2}$. From these expressions, it follows that the tangent of the angle $P\widehat{M}b$ is given by $\frac{m}{n} = \frac{c^2dydx \mp udx\sqrt{c^2dx^2+c^2dy^2-u^2dx^2}}{u^2dx^2-c^2dy^2}$. Then, the boat holding this angle will advance along the given curve $AMC$. *Q.E.I.*

*Corollary 1.* The angle $P\widehat{M}b$ whose tangent is $\frac{m}{n}$, can be conveniently resolved into two angles of which the tangent of one angle is $\frac{dy}{dx}$ and the tangent of the other angle is, indeed, $\frac{\mp\sqrt{c^2dx^2+c^2dy^2-u^2dx^2}}{udx}$. Wherefore, the angle $P\widehat{M}b = \tan^{-1}\frac{dy}{dx} \mp \tan^{-1}\frac{\sqrt{c^2dx^2+c^2dy^2-u^2dx^2}}{udx}$.[i]

*Corollary 2.* Everywhere, therefore, the boat will be able to attain a double angle $P\widehat{M}b$ from which it may follow a given curve $AMC$, provided that $c^2dx^2 + c^2dy^2 > u^2dx^2$ or $c^2ds^2 > u^2dx^2$. For if it happens that somewhere $c^2ds^2 < u^2dx^2$, then the proposed curve cannot be followed by the boat at all.

*Corollary 3.* Therefore, when the proposed curve is describable, it should be so provided that everywhere $\frac{c}{u} > \frac{dx}{ds}$, that is, everywhere $\frac{AD}{PQ} > \frac{Mn}{Mm}$.

*Corollary 4.* Since it is realized that nowhere $Mn > Mm$, if no ordinate of the curve $AMC$ is greater than $AD$, then every curve $AMC$ can be followed by the boat, because in this case it cannot happen that anywhere $\frac{AD}{PQ} < \frac{Mn}{Mm}$.

*Corollary 5.* However, when the double angle of the direction holds, from which a given curve that is possible to be described is known, then, it should be observed that both locations can be found, when each [angle] is positive, if, indeed, the beginning of the motion is placed at $A$. Therefore, a double angle will take place, if $\tan^{-1}\frac{dy}{dx} > \tan^{-1}\frac{\sqrt{c^2dx^2+c^2dy^2-u^2dx^2}}{udx}$,that is, if $udy > \sqrt{c^2dx^2+c^2dy^2-u^2dx^2}$ or $u > c$.[j]

*Corollary 6.* Since the actual velocity of the boat in $M$ which covers the element $Mm$ is $= \frac{cmds}{dx}$, then in our case it will be $= \frac{udy \mp \sqrt{c^2ds^2-u^2dx^2}}{ds}$ , because $dx^2 + dy^2 = ds^2$.

---

[i] The following trigonometric identity was perhaps used in this transformation: $\tan(\propto + \beta) = \frac{\tan\alpha + \tan\beta}{1-\tan\alpha\tan\beta}$, with $\propto = \frac{dy}{dx}$ and $\beta = \frac{\sqrt{c^2dx^2+c^2dy^2-u^2dx^2}}{udx}$, which should result in $(\propto + \beta) = \tan^{-1}\frac{m}{n} = \tan^{-1}\frac{c^2dydx \mp udx\sqrt{c^2dx^2+c^2dy^2-u^2dx^2}}{u^2dx^2-c^2dy^2}$.

[j] See Problem II, Corollary 7: $bf > \sqrt{a^2c^2 + c^2f^2 - a^2b^2}$ or $b > c$, substituting $b$ for $u$ and $f$ for $dy$.

*Corollary 7.* Moreover, the time that it takes to traverse the arc $AM$ will be $= \int \frac{dx}{cm} = \int \frac{ds^2}{udy \mp \sqrt{c^2ds^2 - u^2dx^2}}$. By multiplying the numerator and the denominator of this expression by $udy \mp \sqrt{c^2ds^2 - u^2dx^2}$, this same time will be $= \int \frac{udy \mp \sqrt{c^2ds^2-u^2dx^2}}{u^2-c^2}$.

**Scholion.**

Since the motion of the boat can be faster or slower, from the two angles of direction from which the boat describes a given curve, choose the one whose sine $m$ is greater, because, in this case, it will give the shorter time which is $\int \frac{dx}{cm}$. For this reason, from the above two possible values, I will use the greater one, such that $cm = \frac{udxdy + dx\sqrt{c^2ds^2-u^2dx^2}}{ds^2}$, $cn = \frac{udx^2 - dy\sqrt{c^2ds^2+-u^2dx^2}}{ds^2}$; and the angle $P\widehat{M}b$ will be equal to the sum of the angles, whose tangents are $\frac{dy}{dx}$ and $\frac{\sqrt{c^2ds^2-u^2dx^2}}{udx}$, or the sine of these are $\frac{dy}{ds}$ and $\frac{\sqrt{c^2ds^2-u^2dx^2}}{uds}$. Therefore, the angle $P\widehat{M}b$ will be equal to the sum of the angles, whose cosine are $\frac{dx}{ds}$ and $\frac{udx}{cds}$, whence this angle is easily found. Moreover, the time that the boat takes to traverse the portion $AM$ of the prescribed curve will be $= \int \frac{udy - \sqrt{c^2ds^2-u^2dx^2}}{u^2-c^2}$. Wherefore, if $AM$ is a straight line, so that $y = kx$, the time across $AM$ will be $= \int \frac{kudx - dx\sqrt{c^2+c^2k^2-u^2}}{u^2-c^2}$. It is evident, therefore, that in this case it ought to be $u < c\sqrt{1+k^2}$.

**Problem 5.**

Knowing the velocity of the river at every location, to find the quickest crossing line $AMC$, in which the boat leaving from $A$, sooner reaches $C$, than across any other line joining the points $A$ and $C$.

**Solution.**

Keeping all of what was found in the previous problem, where the nature of the curve $AMC$ was investigated, in which $= \int \frac{udy - \sqrt{c^2ds^2-u^2dx^2}}{u^2-c^2}$ is the formula that gives the minimum value for the time that the boat takes to traverse the arc $AM$ of the curve. Last year I shared a universal method for solving such questions[k], where, if the desired curve in which $\int Zdx$ becomes a maximum or minimum, let $dZ = Fdy + Gdx + Hdp + Idq + Kdr\ etc.$, where $dy = pdx; dp = qdx; dq = rdx\ etc$, then, giving the following equation for the desired curve $0 = Fdx - dH + \frac{d^2I}{dx} - \frac{d^3K}{dx^2}\ etc.$, with $dx$ constant. Then, our formula is reduced to the form $\int Zdx$, by considering, according to the given method, that $dy = pdx$, resulting in $\int Zdx = \int \frac{udy - \sqrt{c^2dx^2+c^2dy^2-u^2dx^2}}{u^2-c^2} = \int \frac{pu - \sqrt{c^2+c^2p^2-u^2}}{u^2-c^2} dx$, or $Z = \frac{pu-\sqrt{c^2+c^2p^2-u^2}}{u^2-c^2}$. Therefore, giving $Z$ as a function of the variables $u$ and $p$ or $x$ and $p$, and, accordingly, $u$ depends on $x$, just as in the problem where I considered that the curve $AQB$ is given. Therefore, we will have that $dZ = Gdx + Hdp$, by considering that the other terms vanish, and, thence, the desired curve will result in $dH = 0$, and also $H = Const.$ It is sufficient, therefore, to find the quantity $H$ from $Z$, which will be obtained by differentiating $Z$, by considering only $p$ as variable. To this end, it will be found that $H = \frac{u}{u^2-c^2} - \frac{c^2p}{(u^2-c^2)\sqrt{c^2+c^2p^2-u^2}} = Const. = \frac{1}{g}$, whence results in the following equation for the desired curve $gu\sqrt{c^2+c^2p^2-u^2} - c^2gp = (u^2-c^2)\sqrt{c^2+c^2p^2-u^2}$, which by squaring results in $p = \frac{c^2+gu-u^2}{c\sqrt{(g-u)^2-c^2}}$. Since we have that $dy = pdx$, the desired curve will be expressed as $dy = \frac{(c^2+gu-u^2)dx}{c\sqrt{(g-u)^2-c^2}}$, from which, because the variables are separated from one another, the curve can be constructed by taking the integral, and, then, $y = \int \frac{(c^2+gu-u^2)dx}{c\sqrt{(g-u)^2-c^2}}$, such that $x$ vanishes for $y = 0$. *Q.E.I.*

[k] Problably, Euler is referring to E56 -- *Curvarum maximi minive proprietate gaudientium inventio nova et facilis,* which, according to the information provided by the Euler Archive, it was written in 1736 (whereas E94 was written in 1738).

*Corollary 1.* Therefore, when the curve $AMC$ is found, through which the boat reaches in the shortest time from $A$ to the given point $C$, the arbitrary constant $g$ should be defined in such a way that, with $x = AB = a, y = BC = b$.

*Corollary 2.* Therefore, since the curve $AMC$ has been found, the angular direction $P\hat{M}b$ will be known, which the boat must hold at any point $M$ in which it is moving through the curve that was found. Since we have that $\sqrt{c^2+c^2p^2-u^2} = \frac{c^2gp}{c^2+gu-u^2} = \frac{cg}{\sqrt{(g-u)^2-c^2}}$, then $\sqrt{c^2ds^2-u^2dx^2} = \frac{cgdx}{\sqrt{(g-u)^2-c^2}}$; thence, the tangent of the angle $P\hat{M}b = \frac{-\sqrt{(g-u)^2-c^2}}{c} = \frac{-(c^2+gu-u^2)dx}{c^2dy}$.

*Corollary 3.* Hence, the secant of the angle $P\hat{M}b$ is $\frac{-g+u}{c}$, and its cosine $= \frac{-c}{g+u}$, which is the sine of the angle $b\hat{M}b$. The boat, therefore, always directed at this angle, reaches the location $C$ from $A$ in the shortest time.

*Corollary 4.* If a river is carried everywhere with the same velocity, or if $u$ is constant, then the line of the fastest crossing will become a straight line: then, in this case, the boat advancing in a straight line reaches $M$ from $A$ most quickly.

*Corollary 5.* If it is considered that $g = \infty$, $P\hat{M}b$ will be a right angle; the boat, therefore, always directed normal to the course of the river, shall cross it in the shortest time; moreover, the curve that it will describe will be governed by the equation $y = \int \frac{udx}{c}$. Then, the boat calls at $C$, such that $BC = \frac{area\ AQBA}{AD}$.

*Corollary 6.* But, as far as the constant $g$ is concerned, it is understood that it ought to be taken in such a way that $(g-u)^2$ is greater than $c^2$. Otherwise, it is observed that the curve found is imaginary.

*Corollary 7.* Since $m$ or the sine of the angle of the angle $P\hat{M}b = \frac{\sqrt{(g-u)^2-c^2}}{g-c}$, then, the time that the boat takes to arrive at $M$ from $A = \int \frac{(g-u)dx}{c\sqrt{(g-u)^2-c^2}}$. And this time is the shortest for which the boat can arrive at $M$ from $A$.

---